# How to manipulate nanoparticle morphology with vacancies

Ilia Smirnov [a] [*], Zbigniev Kaszkur [b] [*], Riccardo Ferrando [c]

[a] University of Warsaw, Warsaw, Poland.

[b] Institute of Physical Chemistry, Warsaw, Poland.

[c] University of Genoa, Genoa, Italy.

[*] Correspondence and requests for materials should be addressed to I.S. (email: snowwhiteman42@gmail.com) or Z.K. (email: zkaszkur@ichf.edu.pl)

## Abstract

Stacking defects in noble metal nanoparticles significantly impact their optical, catalytic, and electrical properties. While some mechanisms behind their formation have been studied, the ability to deliberately manipulate nanoparticle bulk morphology remains largely unexplored. In this work, we introduce a pioneering mechanism - vacancy-driven twinning - that enables the transformation of face-centered cubic (fcc) gold into locally hexagonal close-packed (hcp) structures. This innovative approach, demonstrated through computational simulations, facilitates the creation of realistic, randomly multi-twinned nanoparticle models. By employing a recently developed multidomain X-ray diffraction method (MDXRD), we quantitatively assess the degree of twinning. It is a crucial step in transferring theoretical studies into practical applications. Our work aims to develop tools for modifying and controlling the bulk structure of fcc nanoparticles.

## Introduction

Face-centered cubic (fcc) metal nanoparticles (NPs) attracted great attention due to their unique physical and chemical properties. While studying these properties, most researchers consider only a few parameters: size, distribution of sizes, and shape of crystals. The interior of NPs – bulk morphology is often neglected, although it may affect the catalytical [1,2,3,4,5], optical, and electronic properties [6,7,8,9] of NPs.

Usually explored forms of fcc originated nanocrystals are cuboctahedron (CUB), decahedron (DEC), and icosahedron (ICO). These particles are mixtures of fcc segments with various specially ordered twin planes appearing due to coalescence [10,11] or stepwise/random nucleation [12,13,14]. However, none of these mechanisms provide insight into how one can purposely manipulate the morphology.

Existing morphology-manipulation studies focus on the transition of regular fcc crystalline phase into hexagonal-closed packing (hcp): Au [15,16,17,18], Rh [19], Ag [20,21], Ni [22,23], etc. Most of these particles are non-equilibrium-shaped particles with a bulk morphology that is far from the energy minimum state. Surprisingly, regardless of the progress in the field, the mechanism behind the transition of fcc-to-hcp phase/segments is poorly explored. Another aspect is a lack of theory that connects the existence of ideal crystalline CUB, DEC, ICO models, and non-equilibrium-shaped particles.

In our work, we report a new mechanism triggering the formation of stacking defects – vacancy-driven twinning (VDT). Once the concentration of vacancies exceeds the critical value of ~13%, the ideal fcc stack becomes unstable, leading to the appearance of highly disordered NPs.

To characterize the phenomena, we populate raw CUB Au models with vacancies and perform relaxation/ molecular dynamics (MD) simulations with Sutton-Chen (SC) and Gupta potentials often used for a medium-scale modeling of nanometals. Besides randomly cross-twinned models, this approach allowed us to simulate clusters locally with 2-6 layers thick hexagonal-close packed (hcp) phase. Also, some models showed a 2D and 3D cross-twinning pattern of stacking faults. The models depart from the less defected scheme proposed by Warren [24] and enable the crossing of stacking planes, unlike a scheme proposed by Longo & Martorana [25]. To our knowledge, none of the structural features have been previously reported for computational studies.

Our work aims to provide a theoretical description of a new twinning mechanism and its practical application in manipulations of bulk morphology. To verify the proposed concepts in real NP samples, we demonstrate the application of X-ray diffraction (XRD) tools for routine analysis of bulk morphology. The degree of twinning (number of nanodomains) was estimated with a multidomain XRD approach, while NP density and size were analyzed with a small-angle scattering (SAXS). The proposed techniques are simple and easily available alternatives to the 3D transmission electron microscopy technique.

In chemical literature, one can find numerous examples of diffraction patterns of fcc metals showing evident multitwinning marks. As predicted by Warren the 111 and 200 peaks approach each other, ultimately building 'bridge' intensity while decreasing the crystal size [24]. The 111 to 200 intensity ratio exceeds markedly a value estimated for perfect crystals. The multitwinning in fcc metal nanocrystals synthesized via chemical route appears to be a common feature. In spite of this, computer simulations suggest rather easy ordering of condensing metal vapors leaving not much room for burying defects but able to build up faulty stacked faces that can leave defects once met [12]. As the (111) faces are considered slowly growing, the local seeds can quickly form octahedral shapes and the further growth can be sped up by local twinning and condensation in grooves as suggested by WHS theory and its extension by van de Waal [26,27,28]. On the other hand, the low-temperature chemical reduction can cause more serious local disorder with possible vacancies, quickly relaxed to multitwinned forms. However, the vacancy-driven construction of realistic multitwinned models offers much more freedom, ease, and less computation costs.

## Results

### Effect of vacancies on NP Morphology simulations

To populate fcc NPs with vacancies, we randomly delete atoms from initial strain-free models. Then, the obtained models are relaxed or heated up with MD simulation using Sutton-Chen or Gupta potential. Energy minimization leads to highly disordered structures with partially preserved vacancies. In the meantime, MD heating of disordered clusters results in the ordering and diffusion of vacancies to the surface.

To study the disordering of fcc stacking, we populated raw CUBs of various sizes (11 models) with 0 to 22 % vacancies and relaxed them using SC potential (Table 1, Cluster software [29,30]). To characterize the degree of twinning, we calculated XRD patterns of corresponding models and analyzed the number of domains (Num.dom.) with the multidomain XRD (MDXRD) approach [31]. Short comments on the physical meaning of Num.dom., MDXRD and other bulk analysis methods can be found in Supplementary Information 1.

| CUB, atoms | Vacancies [%]; relaxed models | | | | | | | | |
|---|---|---|---|---|---|---|---|---|---|
| | 0 | 8 | 10 | 12 | 14 | 16 | 18 | 20 | 22 |
| 923 | 1.1 | 1.3 | 1.1 | 1.2 | 3.5 | 2.9 | 8.3 | 6.1 | 7.5 |
| 1415 | 1.1 | 1.3 | 1.3 | 1.1 | 4.3 | 3.4 | 3.8 | 7.3 | 6.6 |
| 2057 | 1.1 | 1.1 | 1.1 | 1.2 | 1.2 | 1.5 | 4.0 | 3.7 | 5.3 |
| 2869 | 1.0 | 1.1 | 1.1 | 1.2 | 1.2 | 1.6 | 5.6 | 8.8 | **3.8**[1] |
| 3871 | 1.1 | 1.1 | 1.0 | 1.1 | 1.5 | 4.6 | 2.6 | 4.4 | **7.7**[2] |
| 5083 | 1.0 | 1.1 | 1.1 | 1.1 | 1.9 | 2.9 | 4.3 | 8.8 | 7.8 |
| 6525 | 1.0 | 1.1 | 1.1 | 1.1 | 1.1 | 6.1 | 10.3 | 5.5 | 8.0 |
| 10179 | 1.0 | 1.1 | 1.1 | 1.2 | 1.2 | 6.0 | 7.7 | 12.6 | 5.1 |
| 14993 | 1.0 | 1.1 | 1.2 | 1.3 | 1.3 | 10.1 | **6.5**[3] | **8.9**[4] | 13.0 |
| 21127 | 1.0 | 1.1 | 1.1 | 1.2 | 2.2 | 6.2 | 8.0 | 13.0 | |
| 28741 | 1.0 | 1.2 | 1.2 | 1.3 | 1.3 | 8.4 | 10.1 | 13.0 | |

Number of domain: 1, 3, 4, 6, 8, 10, 11, 13

**Table 1 Number of domains in relaxed gold CUBs saturated with vacancies.** Corresponding calculated XRD patterns of multidomain particles were analyzed using the MDXRD method. This method provided information on the degree of twinning, which is shown as the number of domains: 1 domain (green) to 13 domains (red). More details in Supplementary Information 2.

We noticed that there are several factors affecting vacancy-driven twinning. First of all, a size dependency of VDT. The smaller a cluster, the higher the percentage of atoms on a surface. Therefore, while creating vacancies in small NPs (by random removal of atoms), there will be an increased number of surface voids that do not lead to bulk twinning. Another critical aspect is an effective conversion of vacancies to stacking faults. One can imagine that there are spatial arrangements of vacancies (their combination/sequences) that will affect the initial structure stronger than others (and vice versa). All these features are less influential in large NPs where most vacancies will likely be created in bulk, and their distribution can be considered uniform.

To minimize the randomness in vacancy distribution, we considered averaging the results of Table 1 over different-size CUBs. It allowed us to determine the average critical concentration of vacancies necessary to cause twinning in CUB models (Fig. 1). In the case of randomly distributed vacancies, it equals ~13.1%. However, there might be spatial arrangements of vacancies that can cause twinning at even lower concentrations.

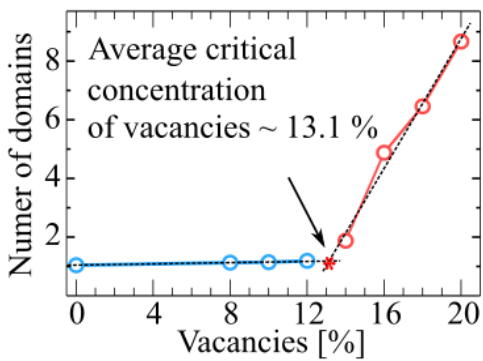

**Fig. 1 The dependence of the number of domains on the amount of vacancies.** Blue points correspond to models with nearly one domain, and pink points represent multiplying twinned particles. The intersection of these lines corresponds to the lowest concentration of vacancies at which twinning is possible.

<span style="color:blue">hcp gold and new stacking patterns</span>

Among the simulated models in Table 1, there are some in which XRD patterns have an additional diffraction peak between 111 and 200 peaks, corresponding to the 101 peaks of hcp Au phase. Fig. 2 shows these hcp-containing clusters superscripted by numbers 1 - 4 in Table 1. It seems that the intensity of the hcp phase peak is dependent not only on the total amount of hcp-phase atoms but also on their distribution and forming multilayer structures. Even though models 3 and 4 (Fig. 2) have similar fractions of hcp atoms, only model 3 has a noticeable additional diffraction peak.

Among all simulated models, the most interesting is CUB 2869 + 22% ( Fig. 2, Model 2, red line with circles). Its crystal structure analysis revealed a 4-layer thick hcp (2H type) domain. By adding two more percent of vacancies (24% in total), the energy minimization led to the appearance of a locally six-layer thick hcp (2H type) segment. This is the first time computational simulation has allowed the local hcp phase to be obtained.

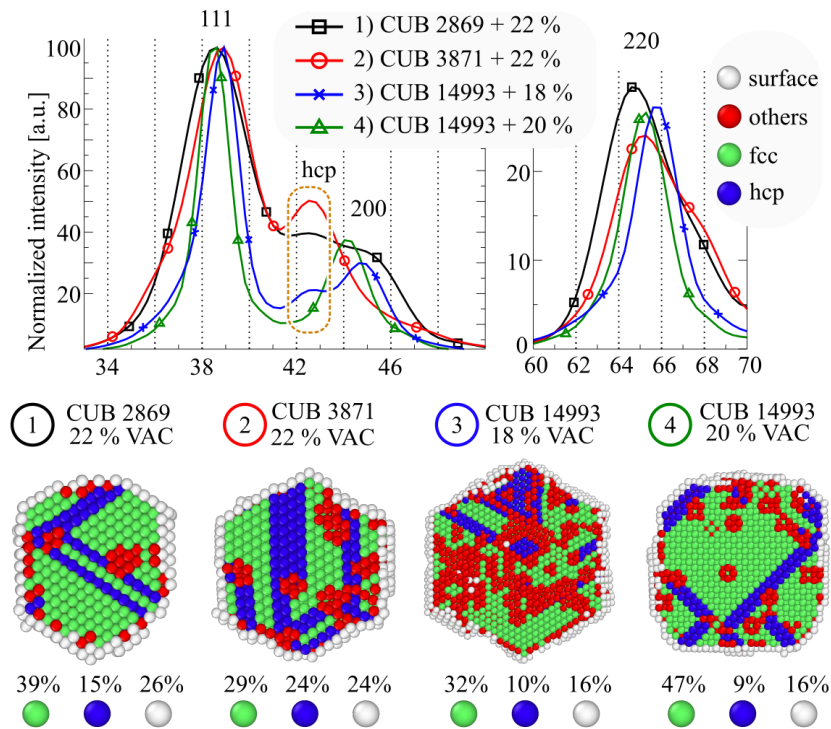

**Fig. 2 Calculated XRD patterns of multidomain models with hcp layers.** The cross-section of the analyzed clusters shows four types of atoms: fcc, hcp, other, and surface. The crystal structure analysis is prepared using OVITO software [32].

However, the obtained 4- and 6-layer thick hcp fragments were not stable upon heating at 293K (Supplementary Fig. 2). These hcp fragments tend to break up into single 2-layer hcp or two 2-layer hcp segments separated by fcc layers (Supplementary Fig. 2). Most likely, it's because these fragments don't penetrate the whole Cluster, making them unstable during heating. This observation highlights the importance of a first step in obtaining randomly twinned NPs. Energy minimization of models with vacancies allows for the simulation of defect–rich but unstable models. Meanwhile, the molecular dynamic produces more stable and ordered structures.

Besides the appearance of a hcp phase, we found a characteristic cross-twinning pattern that was not described in the literature (Fig. 3). This pattern consists of four stacking faults originating from a common center, resulting in 2 dimensional "X" – like pattern. None of the models in Table 1 revealed a stable and well-formed cross-like pattern, so the 4-domain model was made manually (Supplementary Method 1). Ideal one domain CUB (2057 at.) was cut into four pieces; then their positions were adjusted to create four hcp-like stacking planes between domains.

The stability of nanoparticles depends on their potential energy per atom, which is influenced by surface configuration. The 4-domain model, derived from the ideal CUB, has nearly identical surface planes to the mother structure. Despite subtle contributions from stacking and surface defects, the 4-domain Cluster is as stable as the original 1-domain CUB (Supplementary Information 3). MD simulations confirm the stability of the 4-domain structure up to the melting temperature.

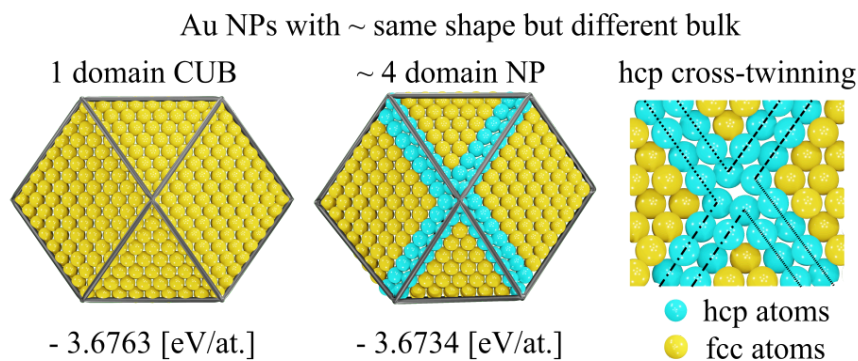

**Fig. 3 1-domain CUB (2057 atoms) and manually created 4-domain Cluster.** All models were relaxed using SC potential. Values below the models correspond to potential energies per atom.

Fig. 3 illustrates the two-dimensional cross-twinning pattern, which is not the only possible pattern. The detailed analysis of models described in Table 1 revealed three-dimensional ones. This cross-twinning has "X"-like pattern (similar to Fig. 3) and lies in three dimensions. In fact, stacking faults coincide with imaginary planes between CUB edges and a central point (24 plains). Similarly to the previous case, energy minimization of software-generated models did produce a well-ordered cross-twinning pattern. It only indicated such a possibility. We tried to build this model manually, but the final 14–domain structure is unstable and tends to collapse during relaxation (Supplementary Data 1).

### Heating of multidomain particles

The relaxed multidomain particles have multiple bulk and surface defects leading to high potential energy in a system (Supplementary Fig. 3). Applying external energy can increase atom mobility, leading to defect diffusion followed by surface and bulk rearrangement. We heated multidomain NPs using MD simulations and analyzed their changes using X-ray scattering in a wide and small-angle X-ray scattering range.
Since the SAXS technique is sensitive to the shape of particles, we used an ideal fcc ball with 5089 atoms as an initial structure. Then the ball was populated with 0 - 22% of vacancies, relaxed, and slowly heated to the melting temperature. All the results can be found in Supplementary Tables 1-4.
Fig. 4 demonstrates some features of the structural evolution of the ball model with 22% of vacancies, where the top part of the picture (Fig. 4 a) provides a visual illustration of the evolution. After initial relaxation, the cluster structure became significantly distorted, with a substantial number of vacancies preserved (represented by red balls – atoms neighboring vacancies). Consequently, this structure exhibits high potential energy (Fig. 4 c), and only "precursor" areas for future twin planes have been formed. Following MD simulation at 300 K, these regions undergo ordering: two new stacking faults formed, and plane faults ordering is in progress. As the temperature increases from 300 K to 550 K, the vacancies diffuse toward the surface (Fig. 4 a; depicted by a sequence of black circles with arrows).

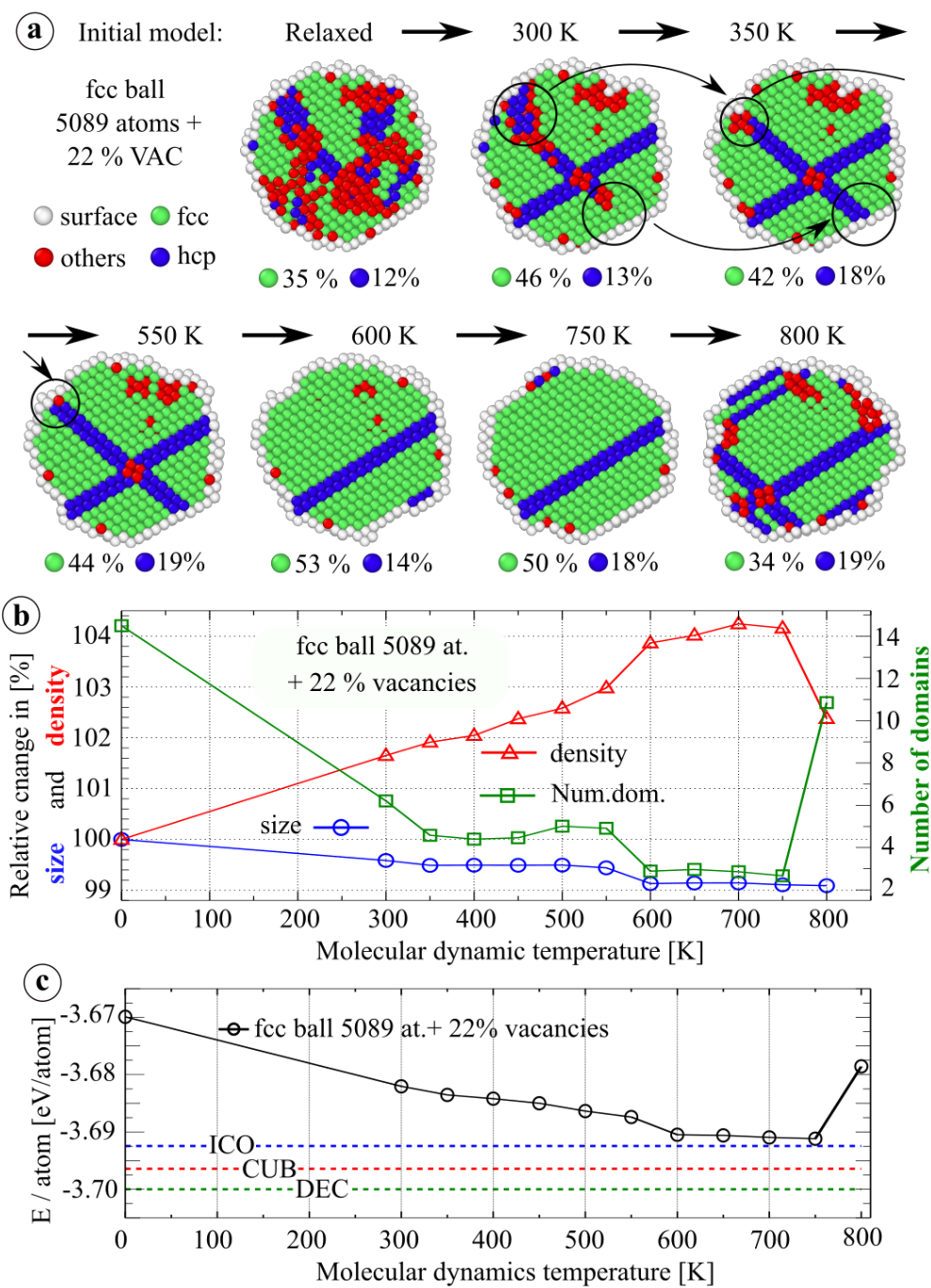

**Fig. 4 Evolution of multidomain NP structure with a temperature. a** Cross sections of fcc ball models with 22% vacancies after relaxation and subsequent MD simulations (SC potential). The duration of each MD step was no less than 50 psec, and after each heating, the clusters were relaxed. **b** The left Y-axis shows relative changes in density (red triangles) and mean size (black circles) calculated via SAXS. The right Y-axis represents the number of domains (green squares) determined by MDXRD. **c** The graph illustrates changes in potential energy per atom after the heating. Blue, red, and green dotted lines correspond to ICO, CUB, and DEC structures of the same size (Supplementary Data 2).

As vacancy concentration decreases, overall density tends to increase. To validate this assumption for each multidomain model, we analyzed corresponding SAXS patterns within a q range of 0.01–0.5 Å$^{-1}$. Fig. 4 b illustrates how high-temperature treatment increases density and reduces mean size. However, upon reaching the premelting temperature (800 K), the cluster accumulates vacancies again, causing secondary twinning. This observation aligns qualitatively with multiple experimental findings [33], [34] showing that the concentration of vacancies increases across the melting point.

Fig. 4 c) depicts the change in potential energy per atom as a function of MD temperature. Remarkably, across the entire temperature range (0–800 K), the density change (Fig. 4 b) correlates proportionally with potential energy change (Fig. 4 c). Interestingly, the slight increase in the number of domains after 500 K did not affect this trend. We notice that vacancies contribute more significantly to the increase in potential energy than twin planes. When void concentration is low, multidomain structure stability is comparable to regular ICO, DEC, and CUB morphologies.

A comparison of a one-domain CUB and 4-domain structures (Fig. 3, Supplementary Information 3), shows that surface is the most important modifier of the total energy (if there are no vacancies inside). The potential energy per atom analysis neglects bulk morphology because the most significant energy changes come from not fully coordinated atoms, i.e. from those neighboring voids and from the surface. However, it might be a critical threshold in the understanding of the formation and growth of NPs. To our knowledge, despite their stability comparable to CUB structures, 4-domain structures have not been observed experimentally. It is likely because of the internal strains created by hcp cross-stacking.

### Comparing results of Sutton-Chen with Gupta potential simulations

Also, we tested the dependency of VDT on the interatomic potentials. The previously obtained results (using SC potential) were compared with simulations based on Gupta potential. Despite some quantitative differences, all trends remain the same. The CUB ( 2869 atoms) saturated with 8% of vacancies showed no twinning and was stable. In the case of 22% of vacancies, the raw CUB structure transformed into a multiply twinned one. The number of domains was similar in both models: relaxed with SC and with Gupta potentials (Supplementary Fig. 4).

To test how interatomic potential affects the temperature-induced ordering, we repeated the simulation from Fig. 4. We headed the same prerelaxed model (fcc ball 5089 at. + 22% of vacancies) using Gupta potential. Again, despite some qualitative differences, all trends remain the same. The higher the temperature, the more ordered the structure (Supplementary Fig. 5). Close to the melting temperature, some vacancies penetrated the bulk and caused secondary twinning. However, the melting temperature obtained by SC and Gupta potentials is different, ~800 and 850 K, respectively. This feature is typical for these potentials.

It seems that both potentials provide qualitatively similar results. However, it is difficult to compare exactly SC and Gupta's potentials in VDT studies as both approaches explore energy relaxation in different ways. During energy minimization, the models with vacancies suffer a series of quick reconfigurations intertwined with periods of slow relaxation (Supplementary Fig. 6). In the complex energy landscape, different potentials and algorithms (conjugated gradients, Hessian, etc.) can follow slightly different paths arriving at a different local minimum.

Spatial arrangements of vacancies

As shown in Table 1, even when the vacancy concentration is high, strong twinning does not necessarily occur. For instance, CUBs composed of 10179 atoms with 20% and 22% vacancy concentrations result in structures with 12.6 and 5.1 domains, respectively. This indicates that VDT is highly sensitive to the initial positions (distribution) of vacancies within the bulk.
To investigate the effect of vacancy distribution, we conducted a series of relaxation simulations (SC potential) on the raw CUB 2869 structure with 22% of defects. In each simulation, the positions of the vacancies were randomly assigned. A total of twelve independent simulations were performed, and the resulting XRD patterns were analyzed to assess the variations in bulk morphology.
The resulting models exhibited significant differences (Supplementary Fig. 7). Fig. 5 demonstrates the substantial impact that the initial positions of voids (with a constant vacancy concentration) can have on the final structure. The red XRD pattern (model #1) displays distinct, well-separated, and narrow diffraction peaks, indicating weak twinning. In contrast, the blue XRD pattern (model #2) shows a noticeable overlap of the 111 and 200 peaks and a broadened 220 peak, suggesting moderate twinning. The black XRD pattern (model #3) features a pronounced overlap of the 111 and 200 peaks, along with a complex profile for the 220 peak, which suggests that this model is heavily defected.

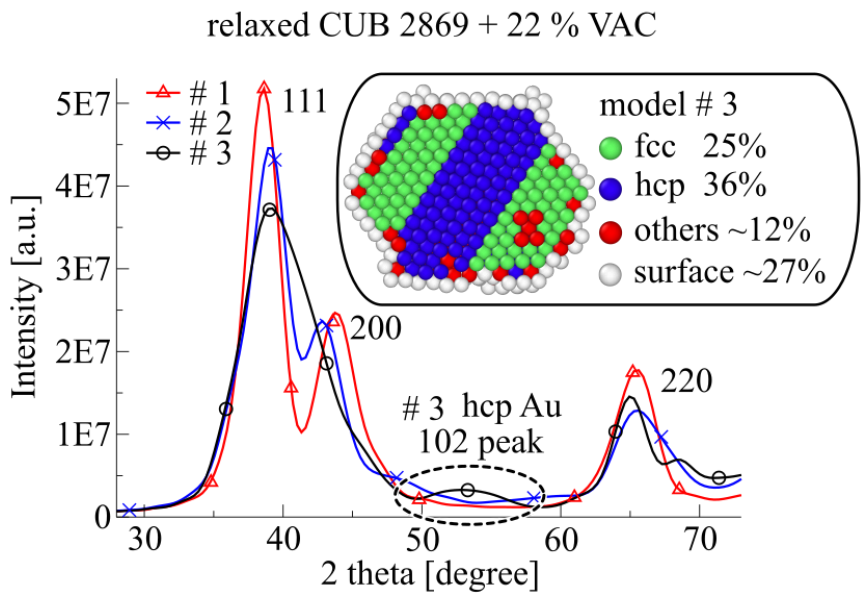

Fig. 5 **Probing different spatial arrangements of vacancies.** The given models illustrate the weakly (red line), moderate (blue line), and strongly (black line) defected structures. The cross-section of the multidomain NP with a 6-layer thick hcp phase corresponds to model #3.

Among the analyzed models, model #3 provides an XRD pattern with an extra XRD peak at ~53.2 degrees that has not been seen for other models. This peak corresponds to the hcp gold phase and appeared because of the 6-layer thick hexagonal phase. Unlike the previously discussed 4-6 thick hcp layers (f.e Fig. 2, model #2), this one penetrates the whole cluster and looks similar to the experimentally observed fcc-2H-fcc gold nanorods[16]. MD simulations (SC potential) showed that model #3 is stable up to the melting temperature. During MD heating, all previously described trends remain valid: as the temperature grows, the preserved vacancies diffuse to the surface, and the stacking layers heal defects and become more ordered.

### VDT in nature: condensation of atoms

The theoretical processes outlined above rely on a top-to-bottom approach when the removal of atoms generates vacancies. However, in reality, it is difficult (not impossible) to imagine how atoms can be extracted from the bulk of NP. Here, we propose two possible scenarios for the appearance of vacancies in real nanoparticles.

The first one is the simple atom condensation model with an intermediate step: the accumulation of vacancies (Supplementary Fig. 8). During condensation, newly deposited atoms may occupy non-optimal positions away from nodal ones. As the number of such defects grows, some voids (vacancies) eventually can be formed.

Theoretical studies suggest that condensation of fcc metals in a vacuum is fast and highly effective. Even if the number of condensed atoms is high, they will likely find a place corresponding to the close packing. However, in reality, in the case of the chemical synthesis route, there are always impurities, solvents, stabilizing agents, etc. Their presence can affect the deposition of atoms, resulting in defect formation.

### VDT in nature: Bimetallic diffusion

Our recent experimental study suggests the appearance of the vacancy-driven twinning mechanism during the phase segregation studies of immiscible AuPt alloys [35]. Alloyed AuPt nanoparticles were heated, which triggered their growth and transformation into Janus particles. The multidomain XRD approach showed that the Pt part of the Janus particle became highly disordered and evolved with temperature, similar to the model presented in Fig. 4. Here, we report a generalized mechanism of the VDT in bimetallic fcc NPs. The detailed description can be found in our work [35].

It is known that macroscopic Au and Pt are immiscible, which is not the case at the nanoscale. Bottom-to-top approaches allow these metals to combine to form well-mixed nanoalloys. Nevertheless, once these nanoparticles grow, they possess more macroscopic-like properties and become unstable. If the temperature is high enough, it triggers the migration of gold atoms from the bulk. Therefore, gold segregates on a surface, forming Janus (or core-shell) NPs.

While gold atoms escape the alloyed phase (Fig. 6), they leave vacancies in the bulk. Since platinum atoms are less mobile than gold, they can not heal the generated vacancies/voids. After the concentration of generated defects exceeds a critical value, the Pt-rich phase becomes unstable and undergoes vacancy-driven twinning. As a result, the experimentally obtained Au-rich and Pt-rich phases consist of ~5 and ~20 domains, respectively. Once the system is heated further, Pt atoms become mobile enough to trigger ordering, and the number of domains decreases.

This case can be used as a guideline for initiating VDT in other fcc samples and synthesizing highly disordered monometallic NPs (Supplementary Method 2). The proposed approach is similar to the Kirkendall effect, wherein during the diffusion process, atoms from the faster-diffusing metal (A) migrate into the slower-diffusing metal (B), creating vacancies within metal A. However, our method differs in two key aspects: the initial state is a homogeneous alloy (AuPt), and vacancies are generated within the metal phase with a lower diffusion rate (Pt). Therefore, our approach is different from the original Kirkendall effect.

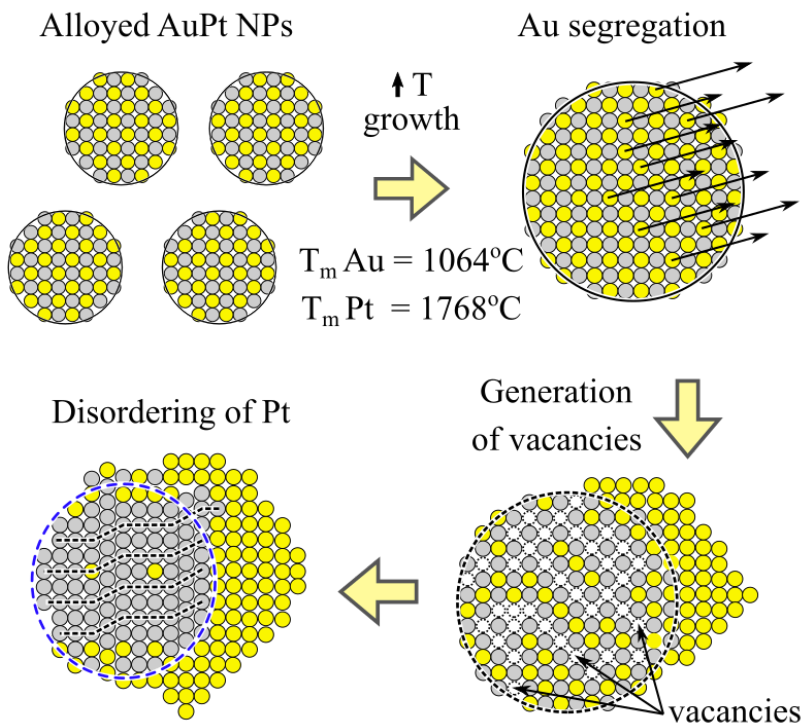

**Fig. 6 Scheme of vacancy-driven twinning in bimetallic NPs during phase segregation.**

In conclusion, we have demonstrated a novel mechanism for the formation of stacking defects in fcc metals. Once the concentration of vacancies exceeds the critical value of approximately 13%, the ideal fcc stack becomes unstable, leading to the appearance of stacking defects. Despite a certain randomness in the appearance of defects, the general trend is as follows: the higher the concentration of vacancies, the more disordered the bulk morphology becomes.

VDT computational simulations allowed us to discover new cross-twinning patterns and obtain a stable 6-layer thick hcp gold phase. Beyond this, the VDT approach offers an alternative avenue for computational studies of NPs. While many existing models rely on regular crystal structures such as CUB, DEC, and ICO, VDT allows the generation of realistic multiple-twinned models with varying degrees of morphology disordering.

Also, we demonstrated a set of tools for analyzing bulk morphology evolution: SAXS and MDXRD. These techniques have enabled us to detect temperature-induced ordering and disordering in Au and AuPt NPs. Our work opens new possibilities for synthesizing NPs with tunable bulk morphologies ranging from highly ordered to disordered ones, which may interest various fields such as optics, catalysis, and electronics. In other words, one can obtain NPs of the same size and shape but with a completely different bulk interior, resulting in different properties.

# Methods

## Computational simulations

### Modeling of NPs structure

Raw "magic number" CUB and fcc ball models were created using Cluster software [29,30]. The same software was used to populate NPs with vacancies.

We used Blender software to manually assemble NP models containing 2D and 3D hcp cross-twinn "X"-like patterns (Fig. 3, Supplementary Method 1). The initial fcc segments were created using Cluster software and then imported into Blender as a set of XYZ coordinates using "PDB/XYZ" add-on. Blender provides a convenient interface for manually editing fcc structures: moving, rotating, shaping, duplicating, etc. The modified structure can then be exported back as a set of XYZ coordinates.

### Energy minimization

The energy minimization implemented in Cluster software utilizes Sutton - Chen potential [36] and was performed until the total energy gradient reached ~10E-6.

Detailed information on energy minimization and molecular dynamics using the Gupta potential can be found in a previous work [37].

### Molecular dynamics

The molecular Dynamics algorithm implemented in Cluster software utilizes SC potential. The simulations shown in Fig. 4 were performed as follows: a raw fcc ball consisting of 5089 atoms was populated with 22% of vacancies and relaxed. Then, the temperature was incrementally increased from 300 K to the melting point at 850 K in steps of 50 K. The duration of each stage was at least 50 ps (in thermostat mode), and the time step used was 0.001 psec. After each heating stage, the corresponding model was relaxed.

Molecular Dynamics simulations (Supplementary Fig. 5) were performed with a Gupta potential. We analyzed the same relaxed fcc ball made of 5089 atoms, which was populated with 22% of vacancies. The temperature incrementally increased from 300 K to 1000 K in steps of 1 K every psec. After each heating stage, the corresponding model was relaxed (using Gupta potential).

### X-ray diffraction data

Powder diffraction patterns (both SAXS and wide-range scattering) were calculated using Cluster software via the Debye summation formula (for Cu radiation). For the analysis of XRD patterns, we used the Pearson type VII function, which is the most commonly used one. SAXS patterns were analyzed in the SASfit program [38], assuming a spherical model with a lognormal size distribution.

### Acknowledgments

This work was supported by the Polish National Science Centre (NCN), grant numbers 2020/39/I/ST5/03385 and 2018/29/B/ST4/00710.

### Corresponding authors

Correspondence to Ilia Smirnov and Zbigniew Kaszkur

### Ethics declarations

#### Competing interests

The authors declare no competing interests.

### Supplementary Information

Supplementary Information is available for this paper.

# References

1. Liu, K. et al. Coherent hexagonal platinum skin on nickel nanocrystals for enhanced hydrogen evolution activity. Nature Communications 14, (2023).
2. Kim, D., Chung, M., Carnis, J. et al. Active site localization of methane oxidation on Pt nanocrystals. Nat Commun 9, 3422 (2018).
3. Dupraz, M., Li, N., Carnis, J. et al. Imaging the facet surface strain state of supported multi-faceted Pt nanoparticles during reaction. Nat Commun 13, 3003 (2022).
4. Nilsson Pingel, T., Jørgensen, M., Yankovich, A.B. et al. Influence of atomic site-specific strain on catalytic activity of supported nanoparticles. Nat Commun 9, 2722 (2018).
5. Carnis, J., Kshirsagar, A.R., Wu, L. et al. Twin boundary migration in an individual platinum nanocrystal during catalytic CO oxidation. Nat Commun 12, 5385 (2021).
6. Liu, X., Luo, J. & Zhu, J. Size effect on the crystal structure of silver nanowires. Nano Letters 6, 408–412 (2006).
7. Park, G. et al. Strain‐Induced Modulation of Localized Surface Plasmon Resonance in Ultrathin Hexagonal Gold Nanoplates. Advanced Materials 33, 2100653 (2021).
8. Qian, X. & Park, H. S. The influence of mechanical strain on the optical properties of spherical gold nanoparticles. Journal of the Mechanics and Physics of Solids 58, 330–345 (2010).
9. Ma, X., Sun, H., Wang, Y., Wu, X. & Zhang, J. Electronic and optical properties of strained noble metals: Implications for applications based on LSPR. Nano Energy 53, 932–939 (2018).
10. Yuk, J. M. et al. High-Resolution EM of Colloidal Nanocrystal Growth Using Graphene Liquid Cells. Science 336, 61–64 (2012).
11. Yuk, J. M. et al. In situ atomic imaging of coalescence of Au nanoparticles on graphene: rotation and grain boundary migration. Chemical Communications 49, 11479 (2013).
12. El, Cesare Roncaglia, Nelli, D., Manuella Cerbelaud & Ferrando, R. Growth mechanisms from tetrahedral seeds to multiply twinned Au nanoparticles revealed by atomistic simulations. Nanoscale horizons 7, 883–889 (2022).
13. Ma, X., Lin, F., Chen, X. & Jin, C. Unveiling Growth Pathways of Multiply Twinned Gold Nanoparticles by In Situ Liquid Cell Transmission Electron Microscopy. ACS Nano 14, 9594–9604 (2020).
14. Xia, Y., Nelli, D., Ferrando, R., Yuan, J. & Li, Z. Y. Shape control of size-selected naked platinum nanocrystals. Nature communications 12, (2021).
15. Huang, X. et al. Synthesis of hexagonal close-packed gold nanostructures. Nature Commun 2, 292 (2011).
16. Fan, Z. et al. Heterophase FCC-2H-FCC gold nanorods. Nature Communications 11, (2020).
17. Huang, X. et al. Graphene oxide‐templated synthesis of ultrathin or tadpole‐shaped au nanowires with alternating hcp and fcc domains. Advanced Materials 24, 979–983 (2012).
18. Fan, Z., Bosman, M., Huang, X. et al. Stabilization of 4H hexagonal phase in gold nanoribbons. Nat Commun 6, 7684 (2015).
19. Duan, H., Yan, N., Yu, R. et al. Ultrathin rhodium nanosheets. Nat Commun 5, 3093 (2014).

20. Liu, X., Luo, J. & Zhu, J. Size effect on the crystal structure of silver nanowires. Nano Letters 6, 408–412 (2006).
21. Chakraborty, I., Shirodkar, S. N., Gohil, S., Waghmare, U. V. & Ayyub, P. A stable, quasi-2d modification of silver: Optical, electronic, vibrational and mechanical properties, and first principles calculations. Journal of Physics: Condensed Matter 26, 025402 (2013).
22. Tzitzios, V. et al. Chemical synthesis and characterization of HCP Ni nanoparticles. Nanotechnology 17, 3750–3755 (2006).
23. Zhuang, Jiahao, et al. "Phase-controlled synthesis of Ni nanocrystals with high catalytic activity in 4-nitrophenol reduction." Journal of Materials Chemistry A 8.42 (2020): 22143-22154.
24. Bertram Eugene Warren. X-ray Diffraction. (Addison-Wesley, 1969).
25. Longo, A. & Martorana, A. Distorted f.c.c. arrangement of gold nanoclusters: a model of spherical particles with microstrains and stacking faults. Journal of Applied Crystallography 41, 446–455 (2008).
26. Wagner, R. S. On the growth of germanium dendrites. Acta metallurgica 8, 57–60 (1960).
27. Hamilton, D. R. & Seidensticker, R. G. Propagation Mechanism of Germanium Dendrites. Journal of Applied Physics 31, 1165–1168 (1960).
28. van de Waal, B. W. Cross-twinning model of fcc crystal growth. Journal of Crystal Growth 158, 153–165 (1996).
29. Mierzwa B. & Kaszkur, Z. Combined XRD-EXAFS software tools for metal nanoclusters. Applied Crystallography. 162-166 (2004).
30. Z.Kaszkur, Program Cluster, https://kaszkur.net.pl/index.php/cluster/.
31. Smirnov, I., Kaszkur, Z. & Hoell, A. Development of nanoparticle bulk morphology analysis: a multidomain XRD approach. Nanoscale 15, 8633–8642 (2023).
32. Stukowski, A. Visualization and analysis of atomistic simulation data with OVITO–the Open Visualization Tool. Modelling and Simulation in Materials Science and Engineering 18, 015012 (2009).
33. Górecki, T. Vacancies and Changes of Physical Properties of Metals at the Melting Point. International Journal of Materials Research 65, 426–431 (1974).
34. Mei, Q. S. & Lu, K. Melting and superheating of crystalline solids: From bulk to nanocrystals. Progress in Materials Science 52, 1175–1262 (2007).
35. Smirnov, Ilia. "Morphology evolution in mono-and bimetallic FCC nanoparticles." (2023).
36. Sutton, A. P. & Chen, J. Long-range Finnis–Sinclair potentials. Philosophical Magazine Letters 61, 139–146 (1990).
37. Rapetti, Daniele, et al. "Machine learning of atomic dynamics and statistical surface identities in gold nanoparticles." Communications Chemistry 6.1 (2023): 143.
38. Breßler, I., Kohlbrecher, J. & Thünemann, A. F. SASfit: a tool for small-angle scattering data analysis using a library of analytical expressions. Journal of Applied Crystallography 48, 1587–1598 (2015).